%% file: main.tex
\documentclass[letterpaper]{article} 
\usepackage{aaai23}  
\usepackage{times}  
\usepackage{helvet}  
\usepackage{courier}  
\usepackage[hyphens]{url}  
\usepackage{graphicx} 
\def\UrlFont{\rm}  
\usepackage{natbib}  
\usepackage{caption} 
\usepackage{algorithm}
\usepackage{algorithmic}

\usepackage{newfloat}
\usepackage{listings}
\usepackage{tikz}
\usepackage{comment}
\usepackage{amsmath,amssymb} 
\usepackage{color}

\usepackage{booktabs}
\usepackage{enumitem}
\usepackage{graphicx}
\usepackage{amsmath}
\usepackage{amssymb}
\usepackage{booktabs}
\usepackage{multirow}

\input{math_command}

\newtheorem{assumption}{Assumption}

\renewcommand\UrlFont\tt\small

\DeclareCaptionStyle{ruled}{labelfont=normalfont,labelsep=colon,strut=off} 
\floatstyle{ruled}
\newfloat{listing}{tb}{lst}{}
\floatname{listing}{Listing}
\title{Deep Equilibrium Models for Video Snapshot Compressive Imaging}
\author{
    Yaping Zhao\textsuperscript{\rm 1,2}, Siming Zheng\textsuperscript{\rm 3,4}, Xin Yuan\textsuperscript{\rm 1,}\thanks{Corresponding author.}
}
\affiliations{
    \textsuperscript{\rm 1} Westlake University, Hangzhou, China\\
    \textsuperscript{\rm 2} The University of Hong Kong, Pokfulam, Hong Kong SAR, China\\
    \textsuperscript{\rm 3} Computer Network Information Center, Chinese Academy of Science, Beijing, China\\
    \textsuperscript{\rm 4} University of Chinese Academy of Sciences, Beijing, China\\
    \textit{\url{zhaoyp@connect.hku.hk}, \url{zhengsiming@cnic.cn}, \url{xyuan@westlake.edu.cn}}

}

\usepackage{bibentry}

\newcommand{\ie}{{\em i.e.}}

\begin{document}

\maketitle

\begin{abstract}
The ability of snapshot compressive imaging (SCI) systems to efficiently capture high-dimensional (HD) data has led to an inverse problem, which consists of recovering the HD signal from the compressed and noisy measurement. While reconstruction algorithms grow fast to solve it with the recent advances of deep learning, the fundamental issue of accurate and \textbf{stable} recovery remains. To this end, we propose deep equilibrium models (DEQ) for video SCI, fusing data-driven regularization and stable convergence in a theoretically sound manner. Each equilibrium model implicitly learns a nonexpansive operator and analytically computes the fixed point, thus enabling unlimited iterative steps and infinite network depth with only a \textbf{constant memory} requirement in training and testing. Specifically, we demonstrate how DEQ can be applied to two existing models for video SCI reconstruction: recurrent neural networks (RNN) and Plug-and-Play (PnP) algorithms. On a variety of datasets and real data, both quantitative and qualitative evaluations of our results demonstrate the {\textbf{effectiveness and stability}} of our proposed method. The code and models are available at: \textcolor{blue}{\url{https://github.com/IndigoPurple/DEQSCI}}.
\end{abstract}

\section{Introduction}
\label{sec:intro}

\begin{figure}[th!]
    \centering
\includegraphics[width=\linewidth]{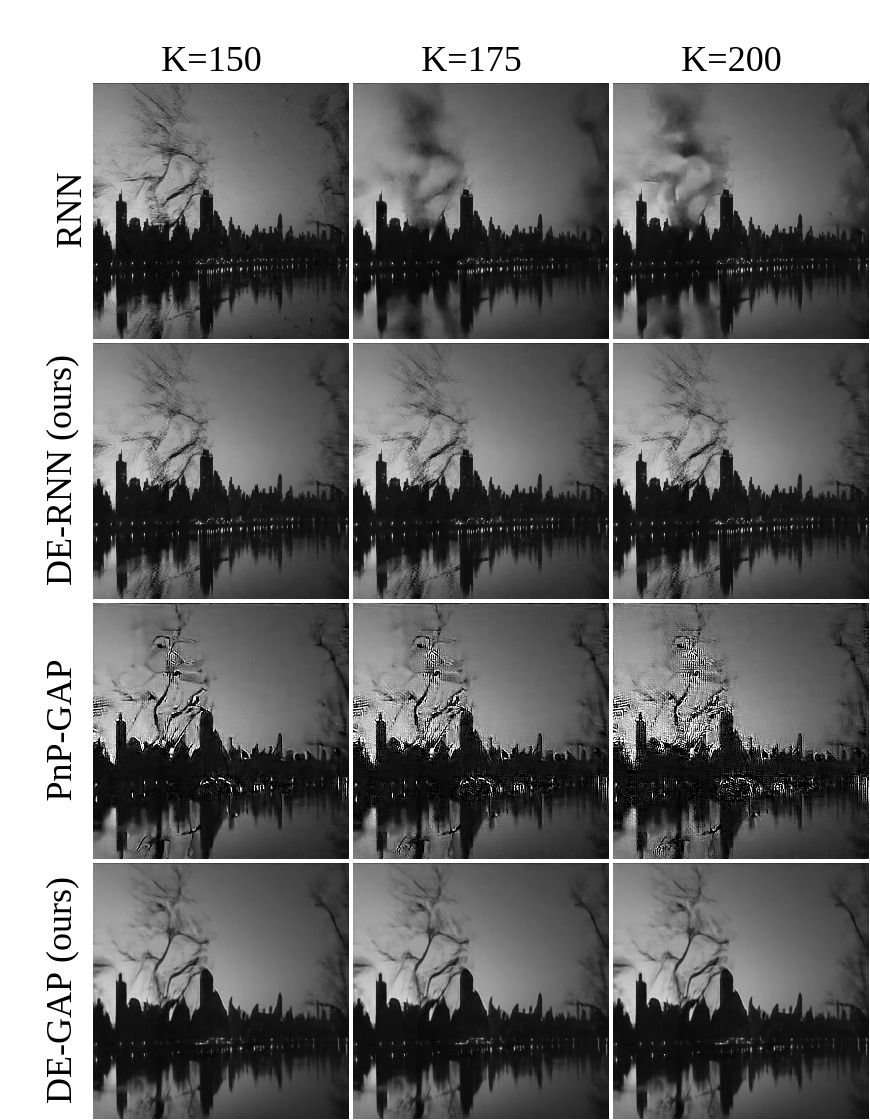}
     \vspace{-5mm}
    \caption{Our proposed deep equilibrium models (DEQ) for SCI can lead to stable recovery as $K$ increases, where $K$ denotes the iteration number during the corresponding optimization progress. We test our model under two different frameworks, \ie, RNN~\cite{cheng2020birnat} and PnP-GAP~\cite{Yuan2020_CVPR_PnP}, the fidelity and stability of our model can be obviously observed. }
     \vspace{-6mm}
    \label{fig:iter_img}
\end{figure}

Aiming at the efficient and effective acquisition of high-dimensional (HD) visual signal, snapshot compressive imaging (SCI) systems have benefited from the advent of novel optical designs to sample the HD data as two-dimensional (2D) measurements. Considering the video SCI system,  the 2D measurement of a video, \textit{i.e.}, a three-dimensional (3D) data-cube leads to an inverse problem. The goal of such an inverse problem is to recover a video from a collection of noisy snapshots, which could be modeled as~\cite{Yuan2021_SPM}:
\begin{align}
    \yv = \Phimat \xv + \ev,
\label{eq:basic}
\end{align}
where $\yv\in {\mathbb R}^n$ is the 2D measurement with $n$ equaling the number of each video frame's pixels, $\Phimat\in {\mathbb R}^{n\times nB}$ is the sensing matrix, $\xv\in {\mathbb R}^{nB}$ is the 3D data (by vectorizing each frame and stacking them), and $\ev$ is the measurement noise; here $B$ denotes that every $B$ video frames are collapsed into a single 2D measurement.
Though algorithms have been fully developed to reconstruct the video from its snapshot measurement in recent years, the fundamental issue remains: this inverse problem is inherently ill-posed, which makes the recovery of the signal $x$ inaccurate and \textbf{unstable} for noise-affected data $\yv$~\cite{jalali2019snapshot}.

The rapid advancement of deep learning and artificial intelligence have empowered a new wave of revolutionary solutions towards these previously intractable problems. For instance, BIRNAT~\cite{cheng2020birnat} employed recurrent neural networks (RNNs) to reconstruct the video frames in a sequential manner and explore the temporal correlation within the video SCI signal. Inspired by particular optimization algorithms,  GAP-net~\cite{meng2020gap}, DUN-3DUnet~\cite{wu2021dense} designed deep unfolding structures, which consist of a fixed number of architecturally identical blocks. The heart of RNN and deep unfolding are deep neural networks, which have posed new challenges due to their ever-growing depth and huge training memory occupation. To overcome these difficulties, inspired by~ \cite{gomez2017reversible}, a recent work (RevSCI)~\cite{cheng2021memory} utilized reversible convolutional neural networks to develop a memory-efficient structure. However, all of these aforementioned algorithms inevitably suffer growing memory occupation with increasing layer depth, and thus models need to be painstakingly designed.

\begin{figure}
    \centering
    \includegraphics[width=\linewidth]{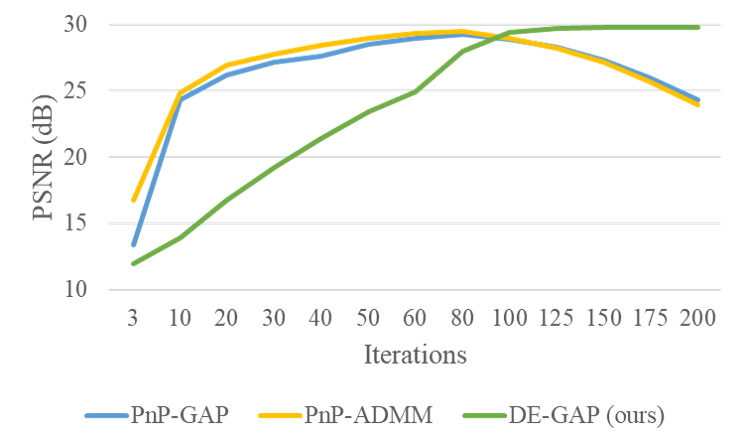}
    \vspace{-6mm}
    \caption{The quantitative comparison of different frameworks with or without our proposed DEQ for SCI. The convergence trends of different algorithms demonstrate that our model's results can converge to a higher level. }
    \vspace{-3mm}
    \label{fig:iter_psnr}
\end{figure}

Inspired by the plug-and-play (PnP) framework~\cite{Venkatakrishnan_13PnP,Sreehari16PnP} which has been proposed for inverse problems with provable convergence~\cite{Chan2017PlugandPlayAF,ryu2019plug}, PnP-FFDNet~\cite{Yuan2020_CVPR_PnP} and PnP-FastDVDNet~\cite{PnP_SCI_TPAMI21} bridged the gap between deep learning and conventional optimization algorithms with the plug-and-play (PnP) framework, utilizing a pre-trained denoiser as the proximal operator. While enjoying the advantages of both data-driven regularization and flexible iterative optimization steps, those algorithms still have hyperparameters to be tuned. Nevertheless, an accurate result must be guaranteed with a proper parameter setting. Due to the intrinsic unstable characteristic of the iterative recovery, even some complicated strategy needs to be employed~\cite{pmlr-v119-wei20b}. As we illustrate in Fig.~\ref{fig:iter_img} and Fig.~\ref{fig:iter_psnr}, the hyperparameters are unavoidable to be handcrafted to achieve satisfactory performance in traditional algorithms.  

An important and interesting research topic in deep learning is to train arbitrary deep networks, in which the deep equilibrium models (DEQ)~\cite{NEURIPS2020_3812f9a5} stands up as the leading method. 
A recent work~\cite{gilton2021deep} leverages DEQ to solve the inverse problems in imaging, which corresponds to the potentially infinite number of iteration steps in the PnP scheme.

To accommodate the state-of-the-art SCI architectures and to enable \textbf{low-memory stable} reconstruction, this paper sets about utilizing DEQ for solving the inverse problem of video SCI. Specifically, we applied DEQ to two existing models for video SCI reconstruction: RNN and PnP. Therefore, the former one is equivalent to an infinite-depth network using only constant memory; the latter one is tuning-free, and directly solves for the fixed point during the iterative optimization process. 
On a variety of simulation and real datasets, quantitative and qualitative evaluations  demonstrate the effectiveness of our proposed method. As shown in Fig.~\ref{fig:iter_psnr}, our reconstruction converges to stable results along with the increasing iterations during optimization. 

In a nutshell, we aim to address the following two challenges which the SCI reconstruction are facing while using deep neural network and iterative optimization algorithms:
\begin{itemize}
    \item How deep should the model be? Can it be infinite?
    \item Is there a tuning-free framework to be used? If yes, how to use it for SCI reconstruction?
\end{itemize}
By employing the most recent development of DEQ, we demonstrate that the answers to all the above questions are positive.
Our specific contributions are as follows:
\begin{itemize}
    \item [1)] We firstly propose deep equilibrium models for video SCI, which fuses data-driven regularization and \textbf{stable} convergence in a theoretically sound manner. 
    \item [2)] Each equilibrium model analytically computes the fixed point, thus enabling unlimited iterative steps and infinite network depth with only a \textbf{constant memory} requirement in training and testing. 
    \item [3)] We analyze convergence for each equilibrium model, to ensure the implicit operators in our models are nonexpansive.
    \item [4)] On a variety of simulations and real datasets, both quantitative and qualitative evaluations of our results demonstrate the {\textbf{effectiveness and stability}} of our proposed method.
\end{itemize}


\section{Related Work}
\label{sec:related}

\subsection{Snapshot Compressive Imaging}
The underlying principle of SCI is to compress the 3D data cube into a 2D measurement by hardware, and then reconstruct the desired signal by algorithms. Considering video SCI, it compresses the spatio-temporal data-cube across the temporal dimension, and thus enables a low-speed camera to capture high-speed scenes. For instance, Llull \textit{et al.}~\cite{llull2013coded} proposed the coded aperture compressive temporal imaging (CACTI) system, which decomposes the 3D cube into its constituent 2D frames and imposes 2D masks for modulation. 

Given the masks and measurements, plenty of algorithms including conventional optimization\cite{liu2018rank,Yang14GMMonline,Yang14GMM,yuan2016generalized}, end-to-end deep learning~\cite{qiao2020deep,zheng2021super, wang2022spatial, cheng2022recurrent, meng2021perception}, deep unfolding~\cite{meng2020gap,wu2021dense} and plug-and-play~\cite{Yuan2020_CVPR_PnP,PnP_SCI_TPAMI21, wu2022adaptive, yang2022revisit} are proposed for reconstruction. To solve the ill-posed problem in Eq.~\eqref{eq:basic}, additional 
regularization is usually needed to ensure accurate and stable recovery with respect to noise perturbation. To this end, these algorithms obtain the estimated value $\hat{\xv}$ of $\xv$ by solving the following problem:
\begin{align}
\label{eq:opt}
    \hat{\xv} =  \mathop{\arg\min}_{ \xv } \frac{1}{2}|| \yv - \Phimat \xv ||_2^2 + R(\xv),
\end{align}
where $|| \yv - \mathbf{\Phi} \xv ||_2^2$ is the fidelity term and $R(\xv)$ is the regularization term.

By introducing an auxiliary parameter $\vv$, the unconstrained optimization in Eq.~\eqref{eq:opt} can be converted into:
\begin{align}
    \label{eq:opt2}
    (\xv, \vv) = \mathop{\arg\min}_{\xv, \vv} \frac{1}{2}|| \yv - \Phi \xv ||_2^2 + R(\vv),~
    s.t.~ \xv=\vv.
\end{align}
Using the alternating direction method of multipliers (ADMM)~\cite{Boyd11ADMM} and introducing another parameter $\uv$, Eq.~\eqref{eq:opt2} could be divided into the following sequence of sub-problems:
\begin{align}
    \xv^{(k+1)} &= \mathop{\arg\min}_{\xv} \frac{1}{2}\| \yv - \Phimat \xv \|_2^2 + \frac{\rho}{2}\|\xv - (\vv^{(k)} - \frac{1}{\rho}\uv^{(k)})\|^2_2, \\
\label{eq:admm-v}
    \vv^{(k+1)} &= \textstyle \mathop{\arg\min}_{\vv} \mu R(\vv) + \frac{\rho}{2}||\vv - (\xv^{(k)} + \frac{1}{\rho}\uv^{(k)}) ||_2^2, \\
    \uv^{(k+1)} &= \textstyle\uv^{(k)} + \rho(\xv^{(k+1)}-\vv^{(k+1)}),
\end{align}
where the superscript $^k$ denotes the iteration number; $\rho$ is the penalty parameter and $\mu$ is the regularization weight.
Since Eq.~\eqref{eq:admm-v} can be regarded as a denoising process of $\vv$, implicitly we have:
\begin{align}
    \vv^{(k+1)} = \textstyle \mathcal{D}^{(k+1)} (\xv^{(k+1)} + \frac{1}{\rho}\uv^{(k)}),
\end{align}
where $\mathcal{D}$ is a denoiser.

On the other hand, generalized alternating projection (GAP)~\cite{Liao14GAP} can be used as a (little bit) lower computational workload algorithm with the following two steps:
\begin{align}
    \xv^{(k+1)} &= \vv^{(k)} + {\Phimat}^\top (\Phimat {\Phimat}^\top)^{-1}(\yv-\Phimat \vv^{(k)}), \label{eq:gap-x}\\
    \vv^{(k+1)} &= \mathcal{D}^{(k+1)} (\xv^{(k+1)}).
\label{eq:gap-v}
\end{align}
Eq.~\eqref{eq:gap-x} can be solved efficiently due to the special structure of $\Phimat$ in SCI~\cite{jalali2019snapshot}.

\subsection{Deep Unfolding}
Inspired by optimization algorithms such as ADMM~\cite{Boyd11ADMM} and GAP~\cite{Liao14GAP}, deep unfolding methods~\cite{ma2019deep,meng2020gap,wu2021dense, yang2022ensemble} are proposed to solve inverse problems in SCI, which consist of a fixed number of architecturally identical blocks. Each of those blocks represents a single iterative step in conventional optimization algorithms. Though deep unfolding successfully assimilate the advantages of the iterative optimization algorithms and could be trained in an end-to-end manner, the fixed number of network blocks in deep unfolding is needed to be kept small for two reasons: $i$) these systems should be concise to keep a high inference speed for real-time reconstruction; $ii$) it is challenging to train deep unfolding networks for numerous stages due to memory limitations.

\subsection{Plug-and-Play}
The latest trend is to bridge the gap between deep learning and optimization with the PnP framework. Yuan \textit{et al.}~\cite{PnP_SCI_TPAMI21} proposed PnP-ADMM framework and PnP-GAP framework, using a pre-trained denoiser as the proximal operator in Eq.~\eqref{eq:admm-v} and Eq.~\eqref{eq:gap-v}, respectively. In contrast to deep unfolding, PnP relieves itself from the limited memory by integrating a flexible denoising module into the iterative optimization process. Nevertheless, it suffers manual parameter tuning in addition to the time-consuming reconstruction process. That is, its performance is highly sensitive to the internal parameter selection, including but not limited to the penalty parameter, the denoising level, and the terminal step number. Moreover, the optimal parameter setting differs image-by-image, depending on the modulation process, noise level, noise type, and the unknown image itself.

\subsection{Memory-Efficient Deep Networks}
Since the important factor that limits the development of deep learning and deep unfolding for SCI is limited memory on hardware devices used for training, to address this issue, RevSCI~\cite{cheng2021memory} developed a memory-efficient network for large-scale video SCI. Using reversible neural networks, where each layer’s input can be calculated from the layer's activation during back-propagation, which means the activation during training is not needed to be stored. Nevertheless,  it still suffers growing memory occupation along with the increasing depth of the network. In contrast, DEQ reduces memory consumption to a constant (\textit{i.e.}, independent of network depth) by directly differentiating through the equilibrium point and thus circumvents the construction and maintenance of layers.
Moreover, DEQ can solve stable estimation, easily extended to larger computing in the test time, while reversible neural networks cannot.

\subsection{Deep Equilibrium Models}
Motivated by the surprisingly recent works~\cite{bai2018trellis,dehghani2018universal,dabre2019recurrent} that employ the same transformation in each layer and still achieve competitive results with the state-of-the-art, Bai \textit{et al.}~\cite{bai2019deep} proposed a new approach to model this process and directly computed the fixed point. To leverage ideas from DEQ, Gilton \textit{et al.}~\cite{gilton2021deep} proposed DEQ for inverse problems in imaging, which corresponds to a potentially infinite number of iteration steps in the PnP reconstruction scheme. 
In this paper, we present a novel approach for video SCI using DEQ, taking both PnP and RNN frameworks into consideration.

\section{Method}
\label{sec:method}
Given measurement $\yv\in {\mathbb R}^{n}$ with compression rate $B$ and sensing matrix $\Phimat\in {\mathbb R}^{n\times nB}$ as input, we consider an optimization iteration or neural network as:
\begin{align}
    \xv^{(k+1)} = f_\theta (\xv^{(k)}; \yv, \Phimat),\quad k = 0,1,\dots, \infty, 
\end{align}
where $\theta$ denotes the weights of embedded neural networks; $\xv^{(k)}\in {\mathbb R}^{nB}$ is the output of the $k^{th}$ iterative step or hidden layer, and $\xv^{(0)} = \mathbf{\Phimat}^\top\yv$; $f_\theta(\cdot\ ; \yv, \Phimat)$ is an iteration map ${\mathbb R}^{nB} \rightarrow {\mathbb R}^{nB}$ towards a stable equilibrium:
\begin{align}
    \lim_{k \rightarrow + \infty} \xv^{(k)} &= \lim_{k \rightarrow + \infty} f_\theta (\xv^{(k)}; \yv, \Phimat) \nonumber\\
    &\equiv \hat{\xv} = f_\theta (\hat{\xv}; \yv, \Phimat),  \label{eq:eq}
\end{align}
where $\hat{\xv}\in {\mathbb R}^{nB}$ denotes the fixed point and reconstruction result.

In this section, we first design different $f_\theta$ for SCI, in terms of the implicit infinite-depth RNN architecture and infinitely iterative PnP framework. Following~\cite{gilton2021deep}, for gradient calculation, we optimize the network wights $\theta$ by approximating the inverse Jacobian. Convergence for specific $f_\theta$ designs is discussed.

\subsection{Forward Pass}
\label{sec:forward}
Unlike the conventional optimization method where the terminal step number is manually chosen or a network where the output is the activation from the limited layers, the result of DEQ is the equilibrium point itself. Therefore, the forward evaluation could be any procedure
that solves for this equilibrium point. Considering SCI reconstruction, we design novel iterative models that converge to equilibrium.

\subsubsection{Recurrent Neural Networks}

\begin{figure}
    \centering
    \includegraphics[width=0.5\linewidth]{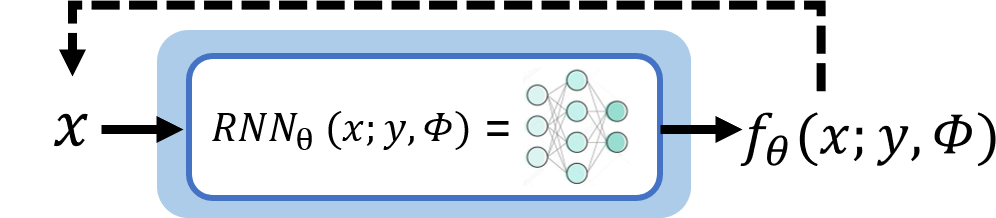}
    \caption{Illustration of our proposed DEQ for SCI using recurrent neural network (RNN), \ie, DE-RNN.}
    \label{fig:pipeline_rnn}
\end{figure}

To achieve integration of DEQ and RNN for video SCI, we have:
\begin{align}
    \xv^{(k+1)} = {\rm RNN}_\theta (\xv^{(k)}, \yv, \Phimat),
\label{eq:dernn-x}
\end{align}
where ${\rm RNN}(\cdot\ )$ is a trainable RNN network learning to iteratively reconstruct effective and stable data. As shown in Fig~\ref{fig:pipeline_rnn}, the corresponding iteration map is:
\begin{align}
    f_\theta (\xv; \yv, \Phimat) = {\rm RNN}_\theta (\xv, \yv, \Phimat).
\label{eq:dernn-f}
\end{align}

\subsubsection{Generalized Alternating Projection}

\begin{figure}
    \centering
    \includegraphics[width=0.7\linewidth]{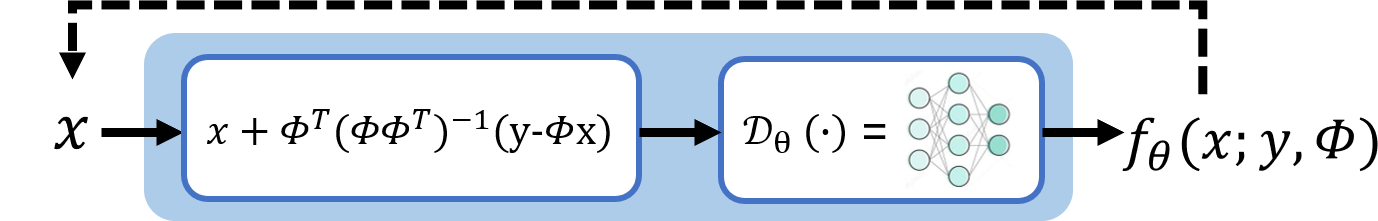}
    \caption{Illustration of our proposed DEQ for SCI using generalized alternating projection (GAP), \ie, DE-GAP.}
    \label{fig:pipeline_degap}
\end{figure}

Regarding the optimization iterations in the GAP method, represented in Eq.~\eqref{eq:gap-x}-\eqref{eq:gap-v}, we iteratively update $\xv$ by: 
\begin{align}
    \xv^{(k+1)} = \textstyle \mathcal{D}^{(k+1)}_\theta \left[\xv^{(k)} + {\Phimat}^\top (\Phimat {\Phimat}^\top)^{-1}(\yv-\Phimat \xv^{(k)}) \right].
\label{eq:degap-x}
\end{align}
Therefore, as illustrated in Fig.~\ref{fig:pipeline_degap}, the iteration map is:
\begin{align}
    f_\theta (\xv; \yv, \Phimat) = \textstyle \mathcal{D}_\theta(\xv + {\Phimat}^\top (\Phimat {\Phimat}^\top)^{-1}(\yv-\Phimat \xv) ).
\label{eq:degap-f}
\end{align}

\subsection{Backward Pass}
\label{sec:backward}
While previous work often utilizes Newton’s method to achieve the equilibrium and then backpropagate through all the Newton iterations, following~\cite{gilton2021deep}, we alternatively adopt another method with high efficiency and constant memory requirement.

\subsubsection{Loss Function}
To optimize network parameters $\theta$, stochastic gradient descent is used to minimize a loss function as follows:
\begin{align}
   \textstyle  \theta^* =   \argmin_\theta \frac{1}{m} \sum_{i=1}^{m} \ell (f_\theta (\hat{\xv}_i; \yv_i, \Phimat_i), \xv^\star_i),  
\end{align}
where $m$ is the number of training samples; $\ell(\cdot, \cdot)$ is a given loss function, $\xv^\star_i$ is the ground truth 3D data of the $i$-{th} training sample, $\yv_i$ is the paired measurement, $\Phimat_i$ denotes the sensing matrix, and $f_\theta (\hat{\xv}_i; \yv_i, \Phimat_i)$ denotes the reconstruction result given as the fixed point $\hat{\xv}$ of the iteration map $f_\theta (\cdot\ ; \yv, \Phimat)$, as derived from Eq.~\eqref{eq:eq}.
The mean-squared error (MSE) loss is used for our video SCI reconstruction:
\begin{align}
    \textstyle  \ell(\hat{\xv}, \xv^\star) = \frac{1}{2} || \hat{\xv} - \xv^\star ||_2^2.
\label{eq:mseloss}
\end{align}
Since the reconstruction result is a fixed point of the iteration map $f_\theta (\cdot\ ; \yv, \Phimat)$, gradient calculation of this loss term could be designed to avoid large memory demand. Following~\cite{gilton2021deep}, we calculate the gradient of the loss term, which takes the network parameters $\theta$ into consideration. 

\subsubsection{Gradient Calculation}
Following~\cite{gilton2021deep}, we calculate the loss gradient. Let $\ell$ be an abbreviation of $\ell(\hat{\xv}, \xv^\star)$ in Eq.~\eqref{eq:mseloss}, then the loss gradient is: 
\begin{align}
\begin{aligned}
\label{eq:gradient}
    \textstyle \frac{\partial \ell}{\partial \theta} &= {\left(\frac{\partial \hat{\xv}}{\partial \theta}\right)}^\top \frac{\partial \ell}{ \partial \hat{\xv}}
    &= {\left(\frac{\partial \hat{\xv}}{\partial \theta}\right)}^\top {\left(\hat{\xv} - \xv^\star \right)},
\end{aligned}
\end{align}
where $\frac{\partial \hat{\xv}}{\partial \theta}$ is the Jacobian of $\hat{\xv}$ evaluated at $\theta$, and $\frac{\partial \ell}{ \partial \hat{\xv}}$ is the gradient of $\ell$ evaluated at $\xv^\star$.

Then to compute the Jacobian $\frac{\partial \hat{\xv}}{\partial \theta}$, we recall the fixed point equation $\hat{\xv} = f_\theta (\hat{\xv}; \yv, \Phimat)$ in Eq.~\eqref{eq:eq}. By implicitly differentiating both sides of this fixed point equation, the Jacobian $\frac{\partial \hat{\xv}}{\partial \theta}$ is solved as:
\begin{align}
   \textstyle  \frac{\partial \hat{\xv}}{\partial \theta} = \left[\boldsymbol{I} - \left.\frac{\partial f_\theta (\xv; \yv, \Phimat)}{\partial \xv}\right|_{\xv = \hat{\xv}}\right]^{-1} \frac{\partial f_\theta (\hat{\xv}; \yv, \Phimat)}{\partial \theta},
\end{align}
which could be plugged into Eq.~\eqref{eq:gradient} and thus get:
\begin{align}
\begin{aligned}
    \textstyle \frac{\partial \ell}{\partial \theta} =   \left[\frac{\partial f_\theta (\hat{\xv}; \yv, \Phimat)}{\partial \theta}\right]^\top \! {\left[\boldsymbol{I} - \left.\frac{\partial f_\theta (\xv; \yv, \Phimat)}{\partial \xv}\right|_{\xv = \hat{\xv}}\right]^{-\top}} 
     {(\hat{\xv} - \xv^\star)},
    \end{aligned}
\end{align}
where $^{-\top}$ denotes the inversion followed by transpose.  
As this method converted gradient calculation to the problem of calculating an inverse Jacobian-vector product, it avoids the backpropagation through many iterations of $f_\theta (\hat{\xv}; \yv, \Phimat)$.  To approximate the inverse Jacobian-vector product, we define the vector $\av^{(\infty)}$ as:
\begin{align}
     \textstyle \av^{(\infty)} = {\left[\boldsymbol{I} - \left.\frac{\partial f_\theta (\xv; \yv, \Phimat)}{\partial \xv}\right|_{\xv = \hat{\xv}}\right]^{-\top}} \! {(\hat{\xv} - \xv^\star)}.
\end{align}
Following~\cite{gilton2021deep}, it is noted that $\av^{(\infty)}$ is a fixed point of the equation:
\begin{align}
\begin{aligned}
\label{eq:jac-eq}
    \av^{(k+1)} =    \textstyle {\left[\left.\frac{\partial f_\theta (\xv; \yv, \Phimat)}{\partial \xv}\right|_{\xv = \hat{\xv}}\right]^{-\top}}\av^{(k)} + {(\hat{\xv} - \xv^\star)},\\ \forall k = 0,1,\dots, \infty.
\end{aligned}
\end{align}
Therefore, the same algorithm used to calculate the fixed point $\hat{\xv}$ could also be used to calculate $\av^{(\infty)}$ . The limit of fixed-point iterations for solving Eq.~\eqref{eq:jac-eq} with initial iterate $\av^{(0)} = \mathbf{0}$ is denoted equivalently to the Neumann series:
\begin{align}
\label{eq:neumann}
    \av^{(\infty)} =  \textstyle \sum_{p=0}^{\infty} 
    \left\{
    {\left[\left.\frac{\partial f_\theta (\xv; \yv, \Phimat)}{\partial \xv}\right|_{\xv = \hat{\xv}}\right]^{\top}}
    \right\}
    ^{p}\!
    {(\hat{\xv} - \xv^\star)}.
\end{align}

To quickly calculate the vector-Jacobian products in Eq.~\eqref{eq:jac-eq} and
Eq.~\eqref{eq:neumann}, a lot of auto-differentiation tools (\textit{e.g.}, autograd packages in Pytorch\cite{paszke2019pytorch}) could be utilized. After the accurate approximation of $\av^{(\infty)}$ is calculated, the gradient in Eq.~\eqref{eq:gradient} is given by:
\begin{align}
     \textstyle\frac{\partial \ell}{\partial \theta} = {\left(\frac{\partial f_\theta (\hat{\xv}; \yv, \Phimat)}{\partial \theta}\right)}^\top \av^{(\infty)}.
\end{align}

\subsection{Convergence Analysis}
\label{sec:convergence}
Given the iteration map $f_\theta(\cdot\ ; \yv, \Phimat): {\mathbb R}^{nB} \rightarrow {\mathbb R}^{nB}$, in this section, we discuss conditions that guarantee the convergence of the proposed deep equilibrium models $\xv^{(k+1)} = f_\theta (\xv^{(k)}; \yv, \Phimat)$ to a fixed-point $\hat{\xv}$ as $k\rightarrow\infty$.   


\begin{assumption}
\textbf{(Convergence of DE-RNN)}.
For all $\xv, \xv' \in {\mathbb R}^{nB}$, if there exists a constant $0\leq c <1$ satisfies that:
\begin{align}
    \| {\rm RNN}_\theta (\xv, \yv, \Phimat) -  {\rm RNN}_\theta (\xv', \yv, \Phimat)\|\leq c\| \xv - \xv' \|,
\label{eq:dernn-conver}
\end{align}
then the DE-RNN iteration map $f_\theta (\xv; \yv, \Phimat)$ is contractive.
\end{assumption}

\begin{assumption}
\textbf{(Convergence of DE-GAP)}.
For all $\xv, \xv' \in {\mathbb R}^{nB}$, if there exists a $\varepsilon > 0$ such that the denoiser $\mathcal{D_\theta}: {\mathbb R}^{nB}\rightarrow{\mathbb R}^{nB}$ satisfies:
\begin{align}
\label{eq:assum}
    \|(\mathcal{D}_\theta - \Imat)(\xv) - (\mathcal{D}_\theta - \Imat)(\xv') \| \leq \varepsilon || \xv - \xv'||,
\end{align}
where $ (\mathcal{D}_\theta - \Imat)(\xv) := \mathcal{D}_\theta(\xv) - \xv$, that is, we assume the map $\mathcal{D}_\theta - \Imat$ is $\varepsilon$-Lipschitz, 
then the DE-GAP iteration map $ f_\theta (\cdot ; \yv, \Phimat) $ defined in Eq.~\eqref{eq:degap-f} satisfies:
\begin{align}
 \textstyle    \| f_\theta (\xv ; \yv, \Phimat) - f_\theta (\xv' ; \yv, \Phimat)\| \leq \eta \| \xv - \xv' \|
\label{eq:degap-conver}
\end{align}
for all $\xv, \xv' \in {\mathbb R}^{nB}$. The coefficient $\eta$ is less than 1, in which case the DE-GAP iteration map $f_\theta (\xv; \yv, \Phimat)$ is contractive.
\end{assumption}

Following~\cite{gilton2021deep}, to prove $f_\theta(\cdot ; \yv, \Phimat)$ is contractive it suffices to show $|| \partial_{\xv} f_\theta(\xv; \yv, \Phimat)|| < 1 $ for all $\xv \in {\mathbb R}^{nB}$, where $|| \cdot ||$ denotes the spectral norm, $\partial_{\xv} f_\theta(\xv; \yv, \Phimat)$ is the Jacobian of $f_\theta(\xv; \yv, \Phimat)$ with respect to $\xv \in {\mathbb R}^{nB}$ given by:
\begin{align}
    \partial_{\xv} f_\theta(\xv; \yv, \Phimat) &= \partial_{\xv} \mathcal{D_\theta}(\xv) (\Imat - {\Phimat}^\top (\Phimat {\Phimat}^\top)^{-1}\Phimat),
\end{align}
where $\partial_{\xv} \mathcal{D_\theta} \in {\mathbb R}^{nB\times nB}$ is the Jacobian of $\mathcal{D_\theta}: {\mathbb R}^{nB}\rightarrow{\mathbb R}^{nB}$ with respect to $\xv \in {\mathbb R}^{nB}$.  

Finally, we derive (details can be found in~\cite{zhao2022mathematical} or supplementary material):
\begin{align}
    || \partial_{\xv} f_\theta(\xv; \yv, \Phimat) ||
    \leq \textstyle (1+\varepsilon) \max_i| 1-\lambda_i |, 
    \label{eq:inequ}
\end{align}
where $\lambda_i$ are eigenvalues of ${\Phimat}^\top (\Phimat {\Phimat}^\top)^{-1}\Phimat $; and the inequality Eq.~\eqref{eq:inequ} is based on the assumption that the map $(\mathcal{D}_\theta - \Imat)(\xv) := \mathcal{D}_\theta(\xv) - \xv$ is $\varepsilon$-Lipschitz.
Therefore the spectral norm of its Jacobian $\partial_{\xv} \mathcal{D_\theta}(\xv) - \Imat$ is bounded by $\eta$, which demonstrates $f_\theta$ is $\eta$-Lipschitz with $\eta = (1+\varepsilon) \max_i| 1-\lambda_i |$. 

It is worth noting that convergence is not yet guaranteed in our calculation above since $\eta$ is larger than 1. In SCI cases, it is challenging to provide a theoretical guarantee. However, we observe our models converge well in the experiments.


\begin{table*}[]
\centering
	\resizebox{1\textwidth}{!}
	{
\begin{tabular}{|c|c|c|c|c|c|c|c|}
\hline
Methods            & Kobe        & Traffic     & Runner      & Drop        & Vehicle     & Aerial      & \textbf{Average}     \\ \hline
GAP-net-AE-S9      & 24.20, 0.570 & 21.13, 0.685 & 29.18, 0.886 & 32.21, 0.907 & 24.19, 0.769 & 24.41, 0.744 & 25.89, 0.760 \\
GAP-TV          & 26.46, 0.885 & 20.89, 0.715 & 28.52, 0.909 & 34.63, 0.970 & 24.82 0.838 & 25.05, 0.828 & 26.73, 0.858 \\
E2E-CNN          & 29.02, 0.861 & 23.45, 0.838 & 34.43, 0.958 & 36.77, 0.974 & 26.40, 0.886 & 27.52, 0.882 & 29.26, 0.900 \\
PnP-FFDnet       & \textbf{30.50}, 0.926 & 24.18, 0.828 & 32.15, 0.933 & \textbf{40.70}, 0.989 & 25.42, 0.849 & 25.27, 0.829 & 29.70, 0.892 \\
\hline
DE-RNN   &       21.46, 0.697      &      19.47, 0.715       &      27.85, 0.818       &   30.16, 0.909          &     23.65, 0.832        &      24.83, 0.855       &      24.53, 0.804      \\
DE-GAP-Unet-3D   &       26.76, 0.866      &   21.42, 0.786          &    30.45, 0.894         &   33.82, 0.963          &     24.94, 0.885       &      24.83, 0.847       &      27.07, 0.878       \\
DE-GAP-RSN-CNN   &       27.33, 0.887      &      22.58, 0.829       &      30.74, 0.903       &   35.95.0.977          &     25.33, 0.899        &      25.57, 0.881       &      27.92, 0.896       \\
DE-GAP-RSN-Unet   &       28.92, 0.939      &      23.68, 0.869       &      32.37, 0.951       &   36.54, 0.972          &     25.50, 0.905        &      25.67, 0.884       &      28.80, 0.913      \\
DE-GAP-CNN   &       28.79, 0.935      &      23.55, 0.864       &      32.35, 0.950       &   38.14, 0.983          &     25.45, 0.903        &      25.84, 0.890       &      29.02, 0.921     \\
DE-GAP-FFDnet   &   29.32, \textbf{0.952}       &      \textbf{24.71}, \textbf{0.907}       &      \textbf{33.06}, \textbf{0.971}       &       39.89, \textbf{0.992}      &       \textbf{25.85}, \textbf{0.905}      &       \textbf{26.02}, \textbf{0.892}      &   \textbf{29.81}, \textbf{0.936}  \\
\hline
\end{tabular}}
\vspace{-3mm}
\caption{ The results in terms of PSNR (dB) and SSIM by different algorithms on classical six datasets for video SCI reconstruction. Compared methods include GAP-net~\cite{meng2020gap}, GAP-TV~\cite{yuan2016generalized}, E2E-CNN~\cite{qiao2020deep} and PnP-FFDnet~\cite{Yuan2020_CVPR_PnP}.}
\label{tab:exp}
\end{table*}

\begin{figure*}
\vspace{-2mm}
    \centering
    \includegraphics[width=1\linewidth]{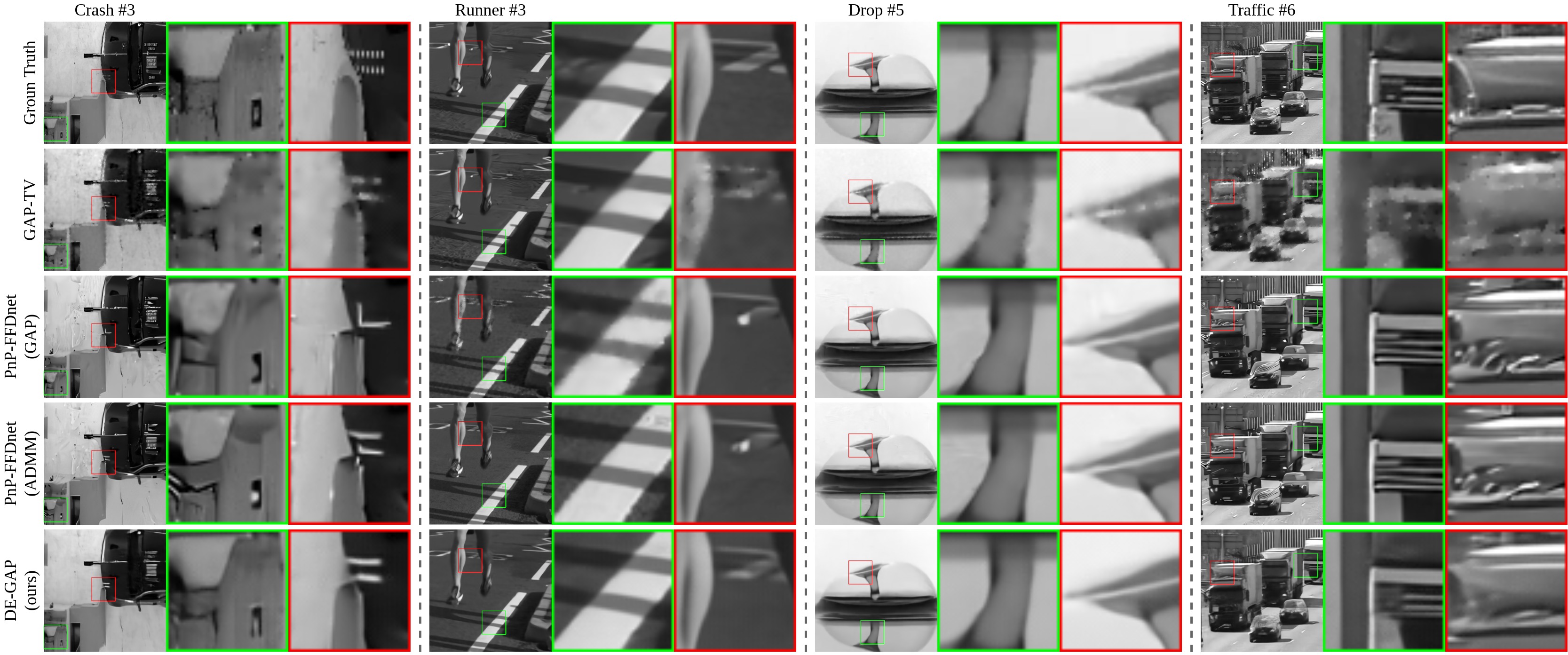}
    \vspace{-7mm}
    \caption{Comparison of selected reconstruction results with the spatial size of $256\times256\times8$. It can be noticed in the zooming areas that GAP-TV is severely blurry, PnP-FFDnet(GAP) and PnP-FFDnet(ADMM) is kind of over smooth around the edges. Our model can achieve cleaner results with sharper edges.}
    \label{fig:results_dataset}
\end{figure*}

\begin{figure*}
\vspace{-2mm}
    \centering
    \includegraphics[width=\linewidth]{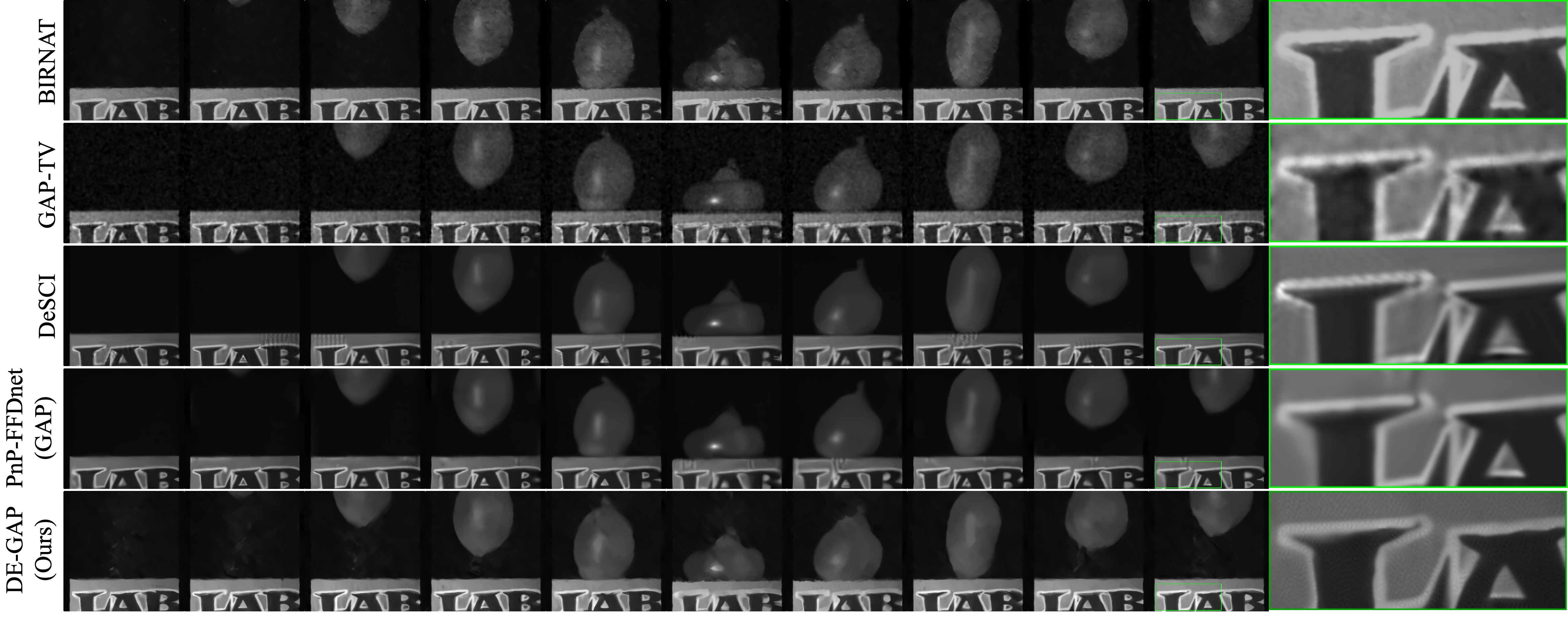}
    \vspace{-7mm}
    \caption{Comparison of selected reconstruction results of real data \textbf{Water Balloon} with the spatial size of $512\times512\times10$. Reconstruction of the real data is more difficult than simulations due to the inevitable measurement noise. As shown in this figure, GAP-TV, DeSCI, and PnP-FFDnet~(GAP) have more artifacts and distortions around margins. Our model can maintain a clear and accurate image structure, thus leading to higher performance.}
    \label{fig:water}
\end{figure*}

\section{Experiment}
\label{sec:exp}

\subsection{Experiment Setting}
\subsubsection{Architecture Specifics}
For our learned network, we have experimented with various network architectures. Specifically, for the DE-RNN model, we adopt the architecture from BIRNAT~\cite{cheng2020birnat}. Regarding its two-stage (forward+backward) RNN as a whole, we iteratively feed the output of the backward RNN back as the input of the forward one. For the DE-GAP model, we employ different neural networks as denoisers $\mathcal{D}_\theta$ and utilize the real spectral norm~\cite{yoshida2017spectral} for convergence purposes. 
We found that some architectures can yield fairly good performance while combining our proposed DEQ for SCI. In a summary, these feasible network architectures are Unet~\cite{ronneberger2015u} with real spectral norm (denoted as RSN-Unet), Unet with 3D convolutional kernels (denoted as Unet-3D), simple CNN networks without and with real spectral norm (denoted as CNN and RSN-CNN, respectively), and FFDnet~\cite{zhang2018ffdnet}.

\subsubsection{Training Details}
Following BIRNAT~\cite{cheng2020birnat}, we choose the dataset DAVIS2017~\cite{pont20172017} for training. DAVIS2017 has $90$ scenes and in total $6208$ frames. We crop its video frames to video patch cubes with the spatial size of $ 256 \times 256 \times 8$, and obtain $26,000$ training samples with data augmentation. Then we train the neural network for $30$ epochs. The initial learning rate is $1\times10^{-3}$ and learning rate decayed is $10\% $ every $10$ epochs. During training, we utilize Anderson acceleration for both the forward and backward pass fixed-point iterations. 

\subsection{Experiment Results}
\subsubsection{Comparisons on Datasets}
For evaluations, we test our proposed DE-RNN and DE-GAP on six classical simulation datasets including \texttt{Kobe}, \texttt{Runner}, \texttt{Drop}, \texttt{Traffic}, \texttt{Vehicle}, and \texttt{Aerial}~\cite{Yuan2020_CVPR_PnP} with the spatial size of ${256\times256}$ and compression ratio $B$=8. Quantitative comparison results with other video SCI reconstruction algorithms including GAP-net~\cite{meng2020gap}, GAP-TV~\cite{yuan2016generalized}, E2E-CNN~\cite{qiao2020deep} and PnP-FFDnet~\cite{Yuan2020_CVPR_PnP} on Peak Signal to Noise Ratio (PSNR) and structured similarity (SSIM)~\cite{wang2004image} are provided in Table~\ref{tab:exp}. 
What stands out in the table is that our method achieves around $0.1$ dB improvement in PSNR and $0.4$ in SSIM. The improvement of SSIM indicates our method could reconstruct images with relative fine structure, which is confirmed by qualitative evaluations in Fig.~\ref{fig:results_dataset}. Specifically, we observe that: i) GAP-TV results have obvious ghosts and fail in high-quality structure reconstruction. For instance, the cars in the \texttt{Traffic} scene reconstructed by GAP-TV are all with heavy blur. ii) in comparison to GAP-TV, our method reconstruct explicit content.  iii) PnP-FFDnet approaches often cause distortion around margins, while our results maintain a clear and accurate structure.

To sum up, 
both the quantitative and qualitative comparisons demonstrate that our method could achieve competitive performance in contrast to other algorithms. We do notice that there are some recent work using complicated deep networks to obtain better results than ours~\cite{cheng2021memory,wu2021dense,wang2021metasci,meng2020gap,zheng2022two}. However, these handcraft designs of different network structures are not necessarily converging to a stable point. By contrast, our paper aims to provide a stable solution for SCI. 

Recalling Fig.~\ref{fig:iter_psnr}, where we have run existing methods and our algorithm for iterations, while RNN and PnP fail in stable recovery, our method could converge to a fixed point and maintain at high-level results. Reconstructed frames in Fig.~\ref{fig:iter_img} further verified this virtue of our proposed algorithm.

\subsubsection{Real-world Data Reconstruction}
We also evaluate the DE-GAP model on real-world dataset \textbf{Water Balloon}~\cite{qiao2020deep} and \textbf{Chopper Wheel}~\cite{llull2013coded}
captured by real video SCI cameras. Note that this is more challenging due to the unavoidable noise inside the real measurements, which demands the high robustness of the algorithm. 

\begin{figure}[h!]
\vspace{-2mm}
  \begin{center}
    \includegraphics[width=0.4\textwidth]{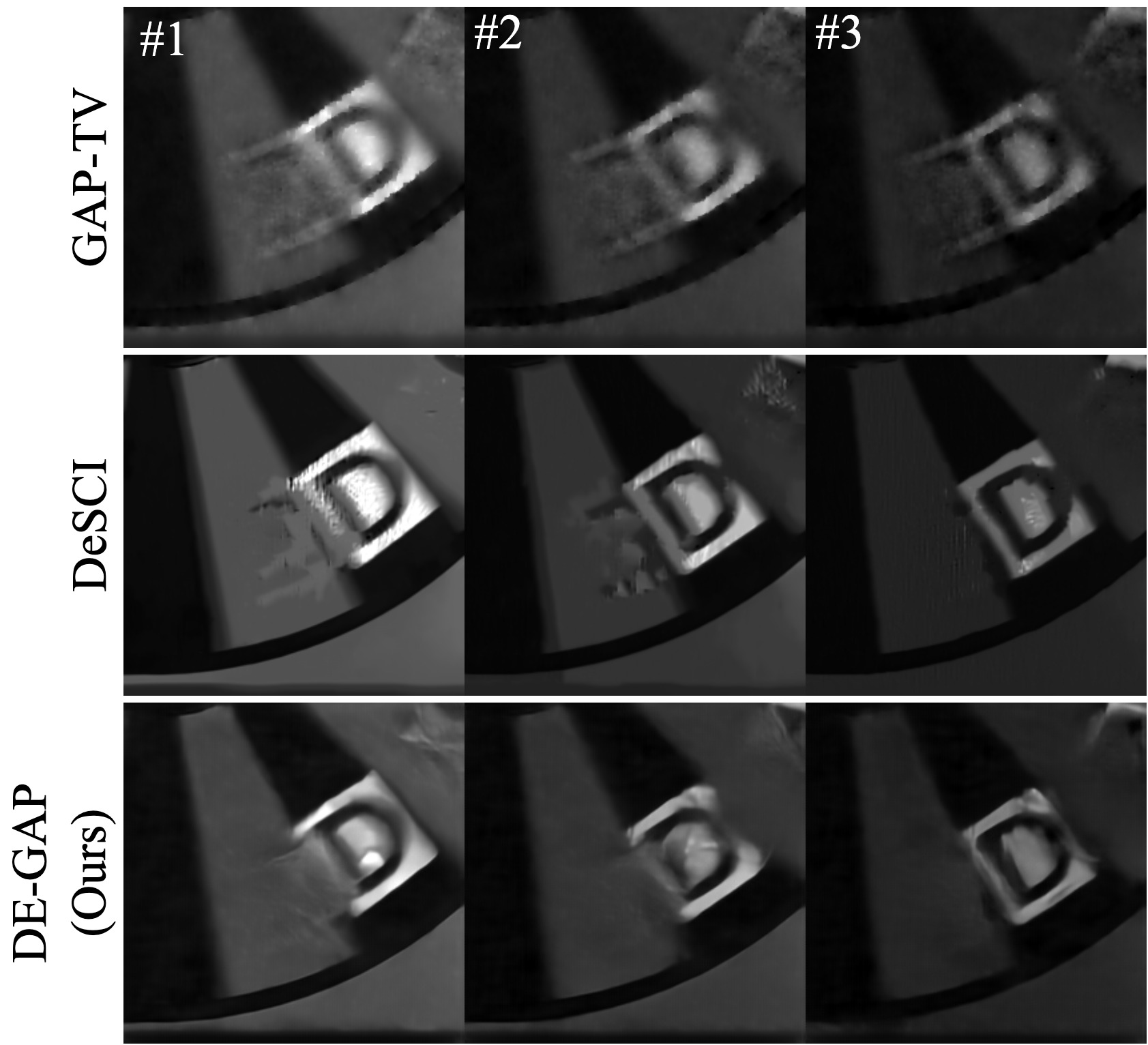}
  \end{center}
  \vspace{-3mm}
  \caption{Comparison of selected reconstruction results of real data \textbf{Chopper Wheel} with the spatial size of $256\times256\times3$. GAP-TV and DeSCI have more ghosts.
  }
  \vspace{-5mm}
    \label{fig:realdata}
\end{figure}

We compare the results with other algorithms including GAP-TV~\cite{yuan2016generalized}, DeSCI~\cite{liu2018rank} and PnP-FFDnet~\cite{Yuan2020_CVPR_PnP}, as shown in Figs.~\ref{fig:water} and \ref{fig:realdata}. The reconstruction results on real-world data demonstrate the effectiveness and generalization of our proposed method. Note that the reconstruction results of real data are achieved by the model trained to utilize the simulation mask, which means that our proposed model is kind of flexible and can achieve stable results by the virtue of the fact that our model can be theoretically infinitely extended. 


Specifically, we observe that:
i) GAP-TV and DeSCI often generate a lot of artifacts and show noisy texture.
ii) PnP-FFDnet has artifacts and distortions around margins.
iii) In contrast to them, our method shows high-quality results with clear content and structure.


\subsubsection{Processing Time}
Though we equivalently realize infinite optimization iterations with deep neural networks plugged in to perform video SCI reconstruction, our designed methods elegantly avoid long inference time. As Table~\ref{tab:runtime} shows, our method needs a relatively short processing time in comparison to other algorithms including GAP-TV~\cite{yuan2016generalized}, DeSCI~\cite{liu2018rank}, PnP-FFDnet~\cite{Yuan2020_CVPR_PnP} and RevSCI~\cite{wang2021metasci}.


\begin{table}[htbp!]
\centering
	\resizebox{.48\textwidth}{!}
	{
\begin{tabular}{|c|c|c|c|c|c|}
\hline
GAP-TV      &   DeSCI   & PnP-FFDnet   & RevSCI & DE-RNN      & DE-GAP    \\ \hline
4.2 & 6180 & 3.0 & 0.19 & 4.68 & 1.90\\
\hline
\end{tabular}}
\vspace{-3mm}
\caption{ Average running time per measurement in seconds by different algorithms on classical six datasets for video SCI reconstruction. 
While permitting unlimited iterative steps and infinite network depth, our method needs a relatively short inference time in contrast to other algorithms.
}
\label{tab:runtime}
\end{table}


\section{Future Work}
Since DEQ under exact gradients may suffer from training time and stability issues, we will incorporate inexact gradients~\cite{geng2021training} and fixed point correction~\cite{bai2022deep} to solve these issues and improve the performance. Our preliminary experiments found that inexact gradients could accelerate the backward passes in training our models by roughly $1.3\sim1.5\times$. Another direction is to integrate DEQ with semantic analysis in SCI~\cite{zhang2022compressive}.

\section{Conclusion}
\label{sec:conclu}
In this paper, to solve the problems of memory requirement and unstable recovery in existing methods, we propose deep equilibrium models for video SCI. Fusing data-driven regularization and stable convergence in a theoretically sound manner, we combine DEQ with existing methods and design two novel models, \textit{i.e.}, DE-RNN and DE-GAP. Each equilibrium model implicitly learns a nonexpansive operator by training the embedded neural network and analytically computes the fixed point, thus enabling unlimited iterative steps and infinite network depth with only a constant memory requirement in the training and inference process. Furthermore, we analyze the convergence conditions for each equilibrium model to ensure the results of our models converge to equilibrium. We evaluate our proposed models using different neural networks as the implicit operator on a variety of simulations and real datasets. In comprehensive comparisons with existing algorithms, both quantitative and qualitative evaluations of our results demonstrate the effectiveness and stability of our proposed method.

\section{Acknowledgements}
This work was supported by the National Natural Science Foundation of China [62271414], Zhejiang Provincial Natural Science Foundation of China [LR23F010001] and Westlake Foundation [2021B1 501-2]. 
The authors would like to thank Research Center for Industries of the Future (RCIF) at Westlake University for supporting this work and the funding from Lochn Optics. The support and funding from Research Postgraduate Student Innovation Award (The University of Hong Kong) is also appreciated.

\bibliography{egbib}

\end{document}

%% file: math_command.tex
\newcommand{\Imat}{{\bf I}}

\newcommand{\av}{\boldsymbol{a}}

\newcommand{\ev}[0]{{\boldsymbol{e}}}

\newcommand{\uv}[0]{{\boldsymbol{u}}}
\newcommand{\vv}{\boldsymbol{v}}

\newcommand{\xv}{\boldsymbol{x}}
\newcommand{\yv}{\boldsymbol{y}}

\newcommand{\Phimat}{\boldsymbol{\Phi}}

\DeclareMathOperator*{\argmin}{arg\,min}